\documentclass[aps,prl,twocolumn,showpacs,showkey,square,numbers,amssymb,amsmath]{revtex4-1}
\usepackage{bm}
\usepackage{times,float}
\usepackage{graphicx}
\usepackage[usenames,dvipsnames,svgnames]{xcolor}
\usepackage{hyperref}
\hypersetup{colorlinks=true, linkcolor=NavyBlue, citecolor=PineGreen,urlcolor=cyan}

\newcommand{\bracket}[1]{\left\langle #1\right\rangle}

\newcommand{\beeq}[1] {\begin{equation}\begin{split}#1\end{split}\end{equation}}

\begin{document}	
\title{Analytic approach for the number statistics of  non-Hermitian random  matrices}
\author{Antonio Tonati\'uh Ramos S\'anchez}
\address{Department of Quantum Physics and Photonics, Institute of Physics, UNAM, P.O. Box 20-364, 01000 Mexico City, Mexico}
\author{Edgar Guzm\'an-Gonz\'alez}
\author{Isaac P\'erez Castillo}
\address{Department of Quantum Physics and Photonics, Institute of Physics, UNAM, P.O. Box 20-364, 01000 Mexico City, Mexico}
\address{London Mathematical Laboratory, 18 Margravine Gardens, London W6 8RH, United Kingdom}
\author{Fernando L. Metz}
\address{Physics Institute, Federal University of Rio Grande do Sul, 91501-970 Porto Alegre, Brazil}
\address{London Mathematical Laboratory, 18 Margravine Gardens, London W6 8RH, United Kingdom}

\begin{abstract}
We introduce a powerful analytic method to study the statistics of the number $\mathcal{N}_{\bm{A}}(\gamma)$ of eigenvalues inside any contour $\gamma \in \mathbb{C}$ for infinitely large non-Hermitian random matrices $\bm{A}$. Our generic approach can be applied to different random matrix ensembles, even when the analytic expression for the joint distribution of eigenvalues is not known. We illustrate the  method on the adjacency matrices of weighted random graphs with asymmetric couplings, for which standard random-matrix tools are inapplicable. The main outcome is an effective theory that determines the cumulant generating function of $\mathcal{N}_{\bm{A}}$ via  a path integral along  $\gamma$, with the path probability distribution following from the solution of a self-consistent equation. We derive the expressions for the mean and the variance of  $\mathcal{N}_{\bm{A}}$ as well as for the rate function governing rare fluctuations of ${\mathcal{N}}_{\bm{A}}{(\gamma)}$. All theoretical results are compared with direct diagonalization of finite random matrices, exhibiting an excellent  agreement.
\end{abstract}

\maketitle

\paragraph{Introduction.}
Since the pioneering work of Wigner \cite{wigner1958distribution}, the study of random matrices has grown into a mature research area, with remarkable applications in  physics, mathematics, biology, statistics, and finance \cite{Mehta,Oxford}. This general character stems mainly from the versatility of random matrix ensembles, which can be thought of as simple but nontrivial models of strongly correlated systems.

The derivation of the joint probability distribution of eigenvalues (JPDE) is one of the most important successes of random matrix theory \cite{Dyson0}, since spectral observables defined in terms of the eigenvalues, including the spectral density and correlation functions \cite{Mehta}, follow from the JPDE. For non-Hermitian random matrices with Gaussian distributed elements, Ginibre deduced the JPDE for matrices with complex and real quaternion entries \cite{Ginibre65}. Due to its simple form, the JPDE for the complex Ginibre case can be mapped in the Boltzmann distribution characterizing an electrostatic system of interacting charges \cite{Ginibre65}. This electrostatic analogy is at the core of the celebrated Dyson's Coulomb fluid approach \cite{Forrester}, where the spectral observables follow from the partition function of an analogous physical system. The situation is considerable more difficult in the Ginibre ensemble with real matrix elements, owing to the existence of a finite fraction of real eigenvalues. In fact, the JPDE for the real Ginibre ensemble was only derived several years after Ginibre's paper in a breakthrough work by Lehmann and Sommers \cite{Lehmann91} (see also \cite{Edelman97,Kanzieper2005}). In this case,  there is no electrostatic analogy and the Coulomb fluid method cannot be applied (see, for insance, \cite{Forrester} for further details).

Among several spectral observables that one may study in random matrix theory, perhaps the most fundamental one is the distribution of the number $\mathcal{N}_D$ of eigenvalues contained in a certain domain $D$, the so-called  {\it number statistics} or {\it full counting statistics} \cite{Mehta}. The study of the fluctuations of $\mathcal{N}_D$ is a rich mathematical problem on itself and, likewise, many problems are transformed into the task of counting how many eigenvalues of a random matrix lie in a certain domain. Examples in this context are the study of the ground state of non-interacting fermions in a harmonic trap \cite{Marino2014,Isaac2014,Marino2016,LacroixA,LacroixB}, the number of stable directions around the stationary points of disordered energy landscapes \cite{Cavagna2000,Wales2003,Majumdar2009}, the number of relevant fluctuation modes in principal component analysis \cite{Vergassola2009,Katzav2010,Majumdar2012}, the localized or extended nature of eigenstates in disordered quantum systems \cite{Mirlin2000,Metz2017,Mirlin2019}, and the stability of large interacting biological systems \cite{Fyodorov2016,Metz2019r}, such as neural
networks \cite{Sommers1988,Rajan2006} and ecosystems \cite{May1972,McMurtrie,allesina2012}.

Thanks to the well-developed machinery of the Coulomb fluid method, a complete picture of the typical and rare fluctuations of $\mathcal{N}_D$ has emerged for Gaussian {\it Hermitian} random matrices with complex, real, and real quaternion entries \cite{Majumdar2009,Marino2014,Isaac2014,Marino2016}. For non-Hermitian random matrices, the question of how many eigenvalues lie outside a disk in the complex plane has been addressed in the case of the real Ginibre ensemble \cite{Allez2014}. However, the number statistics has been fully studied only for the complex Ginibre ensemble \cite{LacroixA,LacroixB}, for which there is an electrostatic analogy for the JPDE and, consequently, the Coulomb fluid method is readily applied. All these studies are limited to circular domains $D$, as more complicated ones are mathematically out of reach. Ironically, non-Hermitian random matrices with {\it real} entries are very relevant for applications, especially in the study of high-dimensional non-equilibrium systems \cite{May1972,McMurtrie,Sompolinsky1988,allesina2012,Sommers1988,Rajan2006,Fyodorov2016,Schuessler2020}, where the matrix entries model the pairwise interactions between the system constituents. Unfortunately, there is no generic analytic method to tackle the number statistics of real asymmetric random matrices and the fluctuations of $\mathcal{N}_D$ remain poorly characterized. 

In this work we design an analytic approach to determine the fluctuations of the number $\mathcal{N}_D$ of eigenvalues inside a domain $D \subset \mathbb{C}$ of arbitrary shape.  We show how the study of the number statistics can be formulated for arbitrary ensembles of infinitely large non-Hermitian random matrices, with real or complex elements, and without relying on the analytic knowledge of the JPDE. In order to exemplify our analytic method, we derive explicit results for the statistics of $\mathcal{N}_D$ in the case of symmetric adjacency matrices of random graphs with asymmetric couplings, for which an analytic expression for the JPDE is not available. The main outcome is a set of effective equations, valid for infinitely large random matrices, which determine all the cumulants and the large deviation function controlling, respectively, the typical and rare fluctuations of $\mathcal{N}_D$. The exactness of our theoretical approach is fully supported by numerical results obtained from the direct diagonalization of finite random matrices. 

\paragraph{The analytic method for the number statistics.}
Let $\lambda_1,...,\lambda_N$ be the eigenvalues of an $N \times N$ non-Hermitian random matrix $\bm{A}$ drawn from a distribution $\mathcal{P}(\bm{A})$. The number of eigenvalues inside a domain $D\subset \mathbb{C}$ enclosed by a contour $\gamma=\partial D$ is given by
\begin{equation}
\mathcal{N}_{\bm{A}}(\gamma) = N \int_{D} d x d y \rho_{\bm{A}} (x,y)\,,  
\label{kla}
\end{equation}  
where $\rho_{\bm{A}}(x,y)$ is the density of eigenvalues around the point $z = x + i y$
\begin{equation}
\rho_{\bm{A}} (x,y) = \frac{1}{N} \sum_{i=1}^N \delta\left(x - {\rm Re} \lambda_i  \right) \delta\left(y - {\rm Im} \lambda_i  \right)\,.
\end{equation}  
In the limit $N \rightarrow \infty$, the statistics of $\mathcal{N}_{\bm{A}}(\gamma)$ is encoded in the cumulant generating function (CGF)
\begin{equation}
\label{eq:cumulat}
  \mathcal{F}_\gamma (\mu)=- \lim _{N\rightarrow \infty} \frac{1}{N}\ln \left\langle e^{-\mu\mathcal{N}_{\bm{A}} (\gamma)} \right\rangle\,,
\end{equation}
with $\langle \dots \rangle$ denoting the ensemble average with the distribution $\mathcal{P}(\bm{A})$. The derivatives of the CGF with respect to $\mu$  determine the cumulants of $\mathcal{N}_{\bm{A}}(\gamma)$. In particular, the intensive mean $N \kappa_1 = \langle \mathcal{N}_{\bm{A}} \rangle$ and variance $N \kappa_2 = \langle \mathcal{N}_{\bm{A}}^2 \rangle - \langle \mathcal{N}_{\bm{A}} \rangle^2 $ read
\begin{equation}
  \kappa_1 = \frac{\partial \mathcal{F}_\gamma (\mu)  }{ \partial \mu} \Bigg{|}_{\mu=0}\,, \qquad
  \kappa_2 = - \frac{\partial^2 \mathcal{F}_\gamma (\mu)  }{ \partial \mu^2} \Bigg{|}_{\mu=0} \,.
  \label{cum}
\end{equation} 
The CGF also provides information about the atypically large fluctuations of $\mathcal{N}_{\bm{A}}$. In fact, the probability that $\mathcal{N}_{\bm{A}} = N n$ ($0 \leq n \leq 1$)
decays, for $N \rightarrow \infty$, as
\begin{equation}
{\rm Prob}_{\gamma}[\mathcal{N}_{\bm{A}} = N n] \asymp  e^{- N \Phi_\gamma(n)}, \label{kalo}
\end{equation}  
where the rate function $\Phi_\gamma(n)$ is determined by the Legendre-Fenchel transform of the CGF \cite{touchette2009large,DemboBook}
\begin{equation}
\Phi_\gamma(n) = - {\rm inf}_{\mu \in \mathbb{R}} \left[\mu n -  \mathcal{F}_\gamma (\mu)  \right]\,.
\end{equation}  
Thus, our goal is precisely to calculate the CGF, since it contains all information about the number statistics.

The first step is to understand how $\mathcal{N}_{\bm{A}}$ depends on $\bm{A}$, so that we can compute, in principle, the ensemble average in Eq.~\eqref{eq:cumulat}. We rewrite the density of eigenvalues as follows \cite{Feinberg1997}
\begin{equation}
\rho_{\bm{A}} (x,y) = \frac{1}{N \pi} \partial_z \partial_{z^{*}} \ln\det{\left[\left(z \bm{I}_N - \bm{A}\right) \left(z \bm{I}_N - \bm{A} \right)^{\dagger}  \right] }\,,
\end{equation}  
where $\partial_z = \frac{1}{2} \left( \frac{\partial}{\partial x} - i  \frac{\partial}{\partial y}  \right)$, $\partial_{z^*} = \frac{1}{2} \left( \frac{\partial}{\partial x} + i  \frac{\partial}{\partial y}  \right)$, and $\bm{I}_N$ is the $N$-dimensional identity matrix. Inserting the above equation back in Eq.~\eqref{kla} and using the Stokes theorem, we obtain
\begin{equation}
\label{eq:N(A)}
\mathcal{N}_{\bm{A}}(\gamma)=-\oint_\gamma     \frac{d z}{2\pi i}  \partial_z \ln Q_{\bm{A}}(z,z^*)\, ,
\end{equation} 
where
\begin{equation}
\label{eq:Q(A)} 
Q_{\bm{A}} (z,z^*)=\frac{1}{\mathrm{det}[(\bm{A}-z \bm{I}_N)(\bm{A}-z \bm{I}_N)^\dagger]}\,,
\end{equation}
with $(\cdots)^*$ and $(\cdots)^\dag$ denoting complex and Hermitian conjugation, respectively. The arbitrary contour $\gamma$ of integration in Eq.~(\ref{eq:N(A)}) is traversed once along the counter-clockwise direction. By discretizing $\gamma$ through a countable set of points $z_1,\dots,z_L$, with $z_{L+1} \equiv z_1$ and $z_{l+1} \equiv z_l + \Delta z_l$, we get the formal identity
\begin{equation}
  \mathcal{N}_{\bm{A}}(\gamma)= - \frac{1}{2 \pi i} \lim_{L \rightarrow \infty} \sum_{l=1}^L \left[ \ln Q_{\bm{A}}(z_{l+1},z_l^*) - \ln Q_{\bm{A}}(z_l,z_l^*) \right],
  \label{klaap}
\end{equation}
and the CGF, Eq.~\eqref{eq:cumulat}, assumes the form
\begin{equation}
\mathcal{F}_\gamma (\mu)= -\frac{1}{N} \ln
\left\langle \prod_{l=1}^L [Q_{\bm{A}} (z_{l+1},z_l^*)]^{ n_{+}  }
             [Q_{\bm{A}}(z_l,z_l^*)]^{ n_{-} } \right\rangle,
             \label{zzpa}
\end{equation}
where $n_{\pm} = \pm \frac{\mu}{2 \pi i}$. The limits $N \rightarrow \infty$ and $L \rightarrow \infty$ are implicit in Eq.~\eqref{zzpa}.

Although Eqs.~\eqref{eq:Q(A)} and \eqref{klaap} expose how $\mathcal{N}_{\bm{A}}$ depends on $\bm{A}$, the calculation of the ensemble average in Eq.~\eqref{zzpa}, with $Q_{\bm{A}}$ in its current form, seems a hopeless task. Using Gaussian integrals, we will rewrite $Q_{\bm{A}}$ in a quadratic form, suitable to compute the average $\langle (\cdots) \rangle$ using methods of statistical physics. Let us introduce the $2 N \times 2 N$ block matrix
\begin{eqnarray}
  \bm{F}_{\eta}(z,z^*)  = \left(\begin{array}{cc} \eta \bm{I}_N & i \left(z \bm{I}_N - \bm{A}  \right) \\
    i \left(z \bm{I}_N - \bm{A}  \right)^{\dagger}  &  \eta \bm{I}_N  \end{array}\right)\,, 
 \end{eqnarray}
which is related to $Q_{\bm{A}}$ via $Q_{\bm{A}} = \lim_{\eta \rightarrow 0^{+}}  \left( \det \bm{F}_{\eta}  \right)^{-1}$. The regularizer $\eta >0$ ensures that $\bm{F}_{\eta}$ has a positive Hermitian part, which enables to represent $Q_{\bm{A}}$ as a Gaussian integral over the spinors $\bm{\psi}_i \in \mathbb{C}^2$ ($i=1,\dots,N$)
\begin{align}
  &Q_{\bm{A}}(z,z^*) = \lim_{\eta \rightarrow 0^{+}} \int \left( \prod_{i=1}^N d \bm{\psi}_i d \bm{\psi}_i^{\dagger}  \right) \nonumber \\
  &\times \exp{\left(-\sum_{i=1}^N \bm{\psi}_i^{\dagger} \bm{M}_\eta (z,z^*) \bm{\psi}_i  + i \sum_{ij=1}^N \bm{\psi}_i^{\dagger} \bm{B}_{ij} \bm{\psi}_j   \right)}\,,
  \label{jja}
\end{align}  
where we introduced the $2 \times 2$ matrices
\begin{eqnarray}
  \bm{M}_\eta (z,z^*) &=& \eta \bm{I}_2 + i \left( z \bm{\sigma}_{+} + z^* \bm{\sigma}_{-}   \right),  \nonumber \\
  \bm{B}_{ij} &=& A_{ij} \bm{\sigma}_{+} + A_{ij}^{\dagger} \bm{\sigma}_{-},
\end{eqnarray}  
and the ladder operators
\begin{eqnarray}
  \bm{\sigma}_{+}  = \left(\begin{array}{cc} 0 &  1 \\
  0  & 0  \end{array}\right)\,, \qquad  \bm{\sigma}_{-}  = \left(\begin{array}{cc} 0 &  0 \\
  1  & 0  \end{array}\right)\,.
 \end{eqnarray}
Equation \eqref{jja} is analogous to the partition function of a system with $N$ spinors placed on the sites of a lattice and coupled through the $2 \times 2$ matrices $\{ \bm{B}_{ij} \}_{i,j=1,\dots,N}$, i.e., $\bm{B}_{ij}$ quantifies the strength of the pairwise interactions between $\bm{\psi}_i$ and $\bm{\psi}_j$. The lattice structure and the distribution of $\{ \bm{B}_{ij} \}_{i,j=1,\dots,N}$ are determined by the specific properties of  $\mathcal{P}(\bm{A})$.

The analogy of $Q_{\bm{A}}$ with a partition function suggests that standard tools of statistical physics can be employed to calculate the CGF. However, there is an additional problem: the presence of the complex-valued exponents $n_{\pm} = \pm \frac{\mu}{2 \pi i}$ in  Eq.~\eqref{zzpa} hampers any direct attempt to evaluate $\langle (\cdots) \rangle$. To overcome this difficulty, we invoke the main strategy of the replica method \cite{Edwards_1975,mezard1987spin} and compute, firstly, the ensemble average in Eq.~\eqref{zzpa} considering $n_\pm \in \mathbb{N}^+$. After performing the limit $N \rightarrow \infty$, the resulting $\mathcal{F}_\gamma (n_{\pm})$ for $n_{\pm} \in \mathbb{N}^+$ is analytically continued to its limiting value as $n_{\pm} \rightarrow \pm \frac{\mu}{2 \pi i}$. Note that the product over $z_1,\dots,z_L$ and the presence of the exponents $n_\pm \in \mathbb{N}^+$ in  Eq.~\eqref{zzpa} do not formally change the quadratic form appearing inside $\langle (\cdots) \rangle$, a feature that is independent of the non-Hermitian random-matrix ensemble under study. Up to this point, the approach is fully general, valid for any contour $\gamma \in \mathbb{C}$ and for arbitrary non-Hermitian random-matrix ensembles, but the success in computing the average $\langle (\cdots) \rangle$ and deriving final equations for the CGF will depend on the choice of $\mathcal{P}(\bm{A})$.

\paragraph{The ensemble of sparse random matrices.}
In this work we illustrate the theory on the adjacency matrix of  random graphs with asymmetric couplings \cite{NewmanBook}. It is convenient to write the matrix elements as $A_{ij} = c_{ij} J_{ij}$, where $c_{ij} \in \{0,1 \}$, $c_{ij} = c_{ji}$, and $c_{ii}=0$. The binary entries $\{ c_{ij} \}_{i,j=1,\dots,N} $ encode the graph structure and $\{ J_{ij} \}_{i,j=1,\dots,N} $ represents the asymmetric interaction strengths, i.e., $J_{ij}$ weights the influence of site $i$ on site $j$. The random variables  $\{ c_{ij} \}_{i,j=1,\dots,N} $ are drawn from 
\begin{eqnarray}
  p_{c}(\{ c_{ij} \}) = \prod_{i < j} \left[\frac{c}{N} \delta_{c_{ij},1} + \left(1 - \frac{c}{N} \right)  \delta_{c_{ij},0}  \right], \label{kll2}
\end{eqnarray}
where $c \in \mathbb{R}^+$ is independent of $N$. Equation \eqref{kll2} yields {\it sparse} random matrices $\bm{A}$ with an average number $c$ of nonzero elements per row and column in the limit $N \rightarrow \infty$. The couplings $\{ J_{ij} \}_{i,j=1,\dots,N} $ are i.i.d.r.v. drawn from a distribution $p_J$. The real asymmetric matrix $\bm{A}$ corresponds to the adjacency matrix
of a weighted random graph with directed edges \cite{NewmanBook}, where the number of neighbors connected to each node follows a Poisson distribution with average $c$ \cite{Erdos59,NewmanBook}. Directed random graphs are key models of networked systems, such as the Internet, neural networks, and food webs (see \cite{BookNetworks} and references therein). The analytic formula for the JPDE of $\bm{A}$ is not known for sparse random-matrix ensembles, which renders traditional tools of random matrix theory unsuitable to study the number statistics.

\paragraph{The effective problem for the CGF.}
In  \footnote{In the Supplemental Information we provide  a detailed account on the analytical derivations}, we explain how one calculates the average $\langle (\cdots) \rangle$ for the random-matrix ensemble given by  Eq.~\eqref{kll2}, takes the limit $N \rightarrow \infty$ through the solution of a saddle-point integral, and finally performs the replica limit $n_{\pm} \rightarrow \pm \frac{\mu}{2 \pi i}$. The main outcome is an \emph{effective theory} defined over the space of functions mapping each point $z$ along the curve $\gamma$ onto a pair of $2\times 2$ matrices $(\bm{\Gamma}(z),\bm{R}(z) )$. The CGF is determined from
\beeq{
\mathcal{F}_{\gamma}(\mu)&=-\frac{c}{2}+\ln\bracket{e^{-\frac{\mu}{2 \pi i} \oint_{\gamma} d z \mathrm{Tr}\left[ \bm{\Gamma}^{-1}(z) \bm{R}(z) \right] } }_{ \{ \bm{\Gamma}, \bm{R} \} } \\
&+\frac{c}{2}\bracket{\bracket{ e^{-\frac{\mu}{2 \pi i} \oint_{\gamma} d z \mathrm{Tr}\left[ \bm{G}(z) \bm{H}(z) \right] } }_{ \bm{J}} }_{ \{ \bm{\Gamma}, \bm{R} \},\{ \bm{\Gamma}^{\prime}, \bm{R}^{\prime} \} }\,,
\label{cumulant2}
}
where we defined the auxiliary $2 \times 2$ matrices at $z \in \gamma$
\beeq{
\bm{G}(z) &= \left[ \bm{I}_2 + \bm{\Gamma}^{\prime}(z) \bm{J} \bm{\Gamma}(z) \bm{J}^{\dagger} \right]^{-1}\,,\\
\bm{H}(z) &=\bm{\Gamma}^{\prime}(z) \bm{J} \bm{R}(z) \bm{J}^{\dagger} + \bm{R}^{\prime}(z) \bm{J} \bm{\Gamma}(z) \bm{J}^{\dagger}\,. 
\label{eq:ffunction}
}
The symbol $\langle \dots \rangle_{\bm{J}}$ stands for the average over 
\begin{equation}
\bm{J} = J \bm{\sigma}_{+} + J^{\prime} \bm{\sigma}_{-},
\end{equation}  
with the real-valued interaction strengths $J$ and $J^{\prime}$ independently drawn from $p_J$. The brackets $\langle (\cdots) \rangle_{ \{ \bm{\Gamma}, \bm{R} \} }$ denote the average over all possible paths $\{ \bm{\Gamma}, \bm{R} \}$ along the curve $\gamma$. For an arbitrary functional  $S[ \{ \bm{\Gamma}, \bm{R} \}]$, we have
\begin{equation}
  \bracket{ S[ \{ \bm{\Gamma}, \bm{R} \} ]}_{ \{ \bm{\Gamma}, \bm{R} \} } = \int  d \{ \bm{\Gamma}, \bm{R} \}  w[ \{ \bm{\Gamma}, \bm{R} \} ]~ S[ \{ \bm{\Gamma}, \bm{R} \}],
\end{equation}
where $w[ \{ \bm{\Gamma}, \bm{R}  \} ]$ is the path probability. A single path $\{ \bm{\Gamma}, \bm{R} \}$ can be thought of as the limit $L \rightarrow \infty$ of a sequence $\{\bm{\Gamma}(z_{l}),\bm{R}(z_{l}) \}_{l=1,\dots,L}$, with $z_l \in \gamma$, while the path integration measure formally reads $d \{ \bm{\Gamma}, \bm{R} \}   = \lim_{L \rightarrow \infty} \prod_{l=1}^L d \bm{\Gamma}(z_{l}) d \bm{R}(z_{l})$. The path probability distribution $w[ \{ \bm{\Gamma}, \bm{R} \} ]$ follows from the solution of the self-consistency equation
\begin{align}
  w[ &\{ \bm{\Gamma}, \bm{R} \}]  = \frac{1}{\Lambda} \sum_{k=0}^\infty\frac{e^{-c}c^k}{k!} \int \left( \prod_{r=1}^k    d \{ \bm{R}_r , \bm{\Gamma}_r \}  w[ \{ \bm{R}_r,\bm{\Gamma}_r \}]  \right) \nonumber \\
&\times e^{\mu  W[ \{ \bm{\Gamma},\bm{R} \}]  }  \bracket{\delta_{({\rm F)}}\left( \bm{R} - \bm{\Pi}_k \right)\delta_{({\rm F})}\left(\bm{\Gamma} - \bm{\chi}_k \right)}_{\bm{J}_{1,\dots,k}} ,
\label{eq:W}	
\end{align}
where $\langle\dots\rangle_{ \bm{J}_{1,\dots,k}  }$ is the average over $\bm{J}_1,\dots,\bm{J}_k$, $\delta_{({\rm F)}}$ represents the functional Dirac  delta in the path space, and $\Lambda$ ensures the normalization of $w[\{ \bm{\Gamma}, \bm{R} \}]$. We have also introduced the $2 \times 2$ matrices at  $z \in \gamma$
\begin{align*}
	\bm{\chi}_k(z) &= \left[ \bm{M}_{\eta}(z,z^\star)+ \sum_{r=1}^k \bm{J}_r \bm{\Gamma}_{r}(z) \bm{J}_r^{\dagger} \right]^{-1}, \\
	\bm{\Pi}_k(z) &=-\bm{\chi}_{k}(z) \left[i \bm{\sigma}_{+} + \sum_{r=1}^{k} \bm{J}_r \bm{R}_r(z) \bm{J}_r^{\dagger} \right] \bm{\chi}_{k}(z).
\end{align*}
The statistical contribution of each path  in Eq.~\eqref{eq:W} is also weighted according to an exponential factor, controlled by
\begin{equation}
W[ \{ \bm{\Gamma},\bm{R} \} ] =\oint_{\gamma} \frac{ d z}{2\pi i} \mathrm{Tr}\left[ \bm{\Gamma}^{-1}(z) \bm{R}(z) \right].
\end{equation}
Equation \eqref{eq:W} must be solved in the limit $\eta \rightarrow 0^+$.

\paragraph{Numerical results.}
Equation \eqref{cumulant2} for the CGF, together with Eq.~\eqref{eq:W} for the path probability, form the main outcome of our work, from which one can study the statistics of $\mathcal{N}_{\bm{A}}$ for $N \rightarrow \infty$. In general, Eq.~\eqref{eq:W} has no explicit solution and therefore one has to resort to a population dynamics approach \cite{Mzard2001,metz2016large} to obtain numerical solutions for $w[ \{ \bm{\Gamma}, \bm{R} \}]$. In this numerical procedure, we discretize a single path over $\gamma$ through a finite set $\{ \bm{\Gamma}(z_i), \bm{R}(z_i) \}_{i=1,\dots,L}$ containing $L$ two-dimensional random matrices that are sampled consistently with Eq.~\eqref{eq:W} via a Monte Carlo scheme. The discrete representation of $w[\{ \bm{\Gamma}, \bm{R} \}]$ as the joint distribution of $\{ \bm{\Gamma}(z_i), \bm{R}(z_i) \}_{i=1,\dots,L}$ does not factorize, because different points along $\gamma$ are correlated through the randomness of the graph ensemble.

\begin{figure}[h]
\centering
\includegraphics[height=5cm,width=8cm]{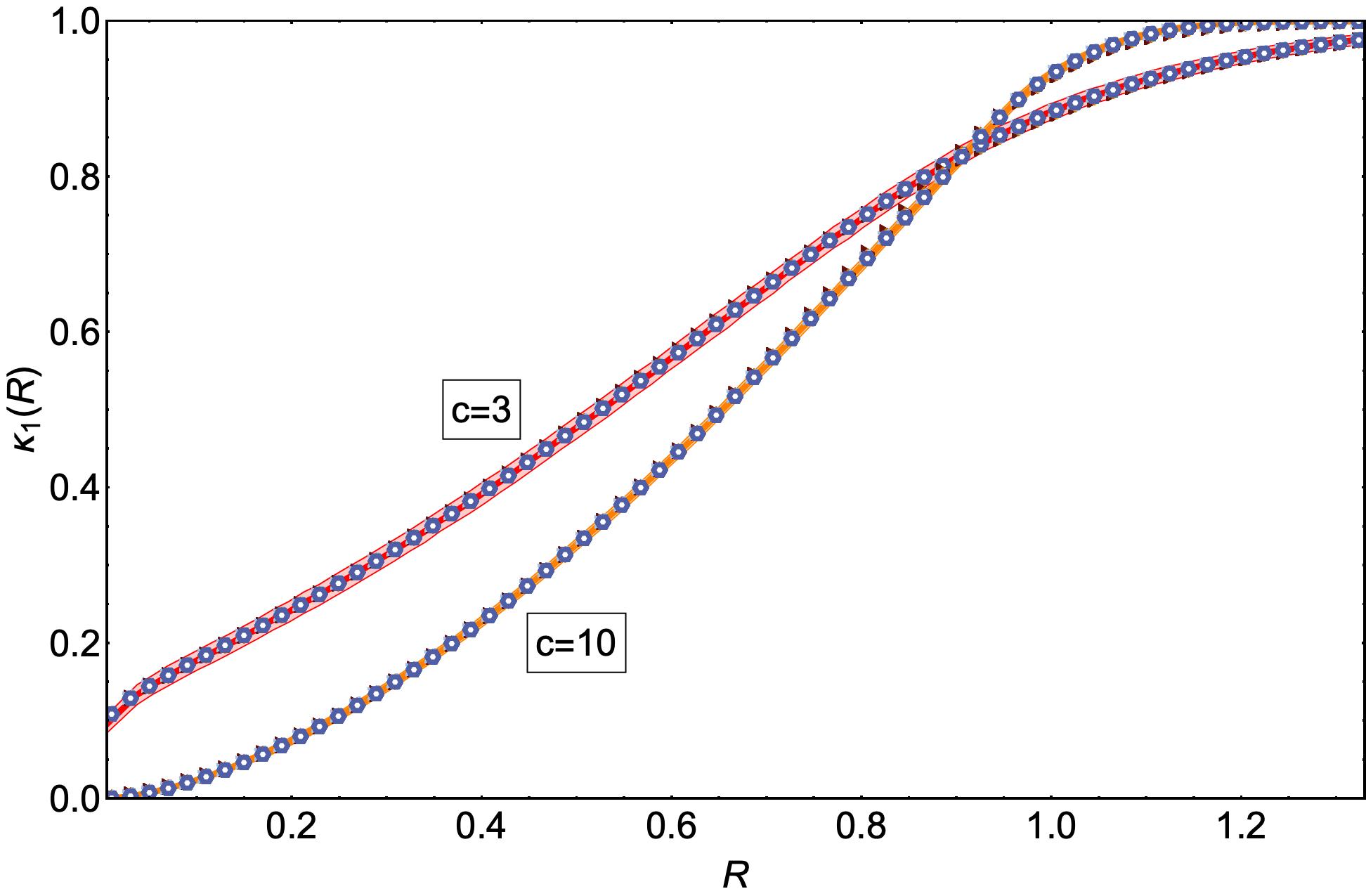}
\includegraphics[height=5cm,width=8cm]{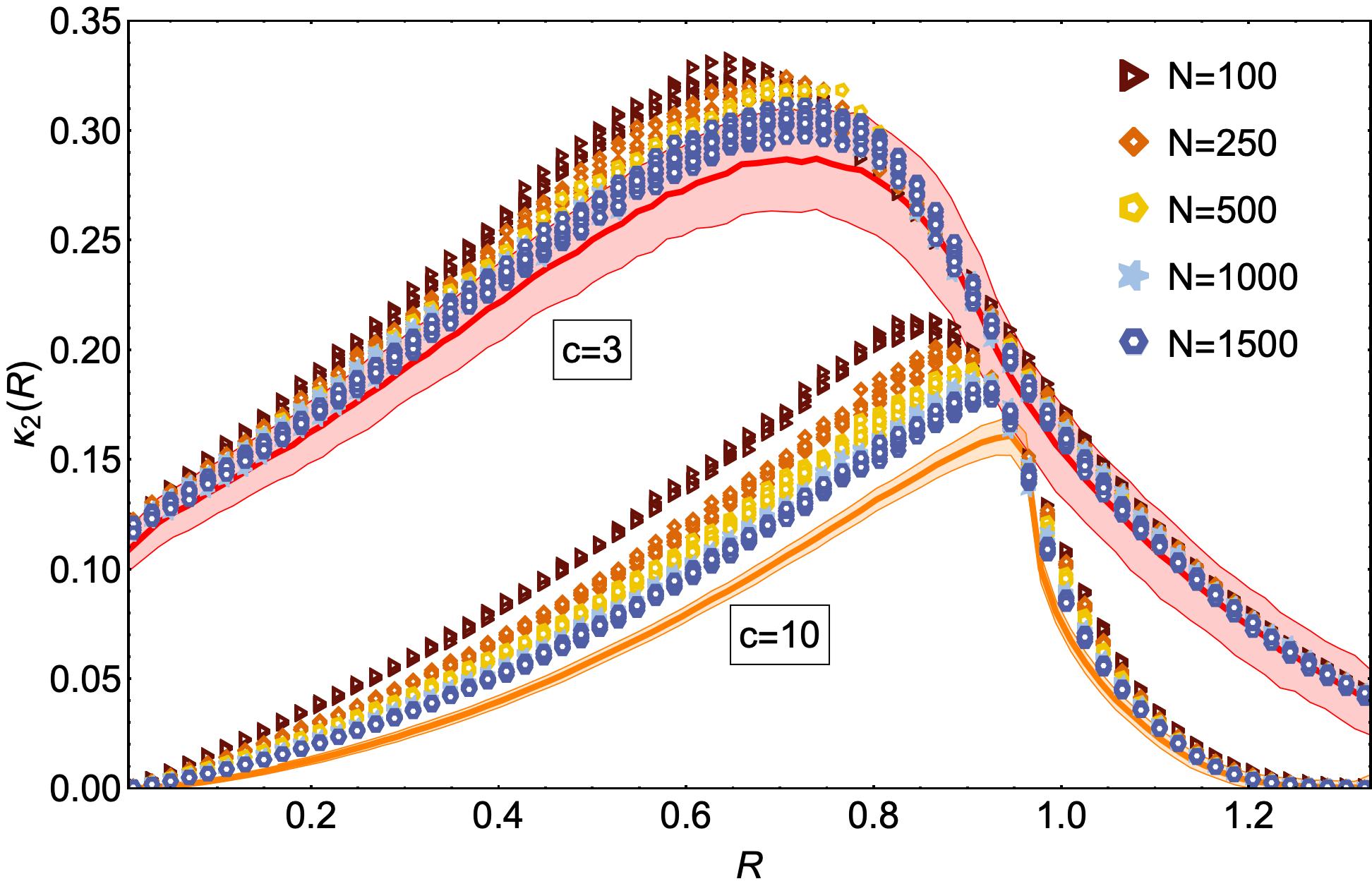}
\caption{The intensive mean $\kappa_1$ and variance $\kappa_2$ of the number of eigenvalues $\mathcal{N}_{\bm{A}}$ within a disk of radius $R$ centered at the origin of the complex plane. The random matrix $\bm{A}$  represents the adjacency matrix of random graphs (see Eq.~(\ref{kll2})) with mean connectivities $c=3$ (red solid line) and $c=10$ (orange solid line). The asymmetric interactions are independently drawn from a Gaussian distribution with zero mean and variance $1/c$. The theoretical solid lines are obtained from the solutions of Eqs.~\eqref{eq:Kumulant1} and \eqref{eq:Kumulant2} using the population dynamics algorithm, while the markers are numerical diagonalization results of $N \times N$ random matrices. The diagonalization results are averaged over $10^4$ samples and the process is repeated $10$ times, yielding the scatter plots shown in the figures. }
\label{fig:firstCumulant}
\end{figure}

Let us present explicit results for the fluctuations of $\mathcal{N}_{\bm{A}}$ and compare our effective theory for $N \rightarrow \infty$ with direct diagonalization of finite random matrices. The mean and the variance of $\mathcal{N}_{\bm{A}}$ follow from Eqs.~(\ref{cum}) 
\begin{eqnarray}
  \kappa_1 &=& -\bracket{  W[ \{ \bm{\Gamma},\bm{R}  \}]  }_{  \{  \bm{\Gamma}, \bm{R}  \}  } \nonumber \\
  &-& \frac{c}{2} \bracket{\oint_\gamma\frac{dz}{2\pi i} \bracket{ \text{Tr}\left[ \bm{G}(z) \bm{H}(z) \right]
    }_{\bm{J} } }_{ \{  \bm{\Gamma}, \bm{R}  \},\{  \bm{\Gamma}^{\prime}, \bm{R}^{\prime}  \}   }\,,
  \label{eq:Kumulant1} \\
  \kappa_2&=& \bracket{   \left( W[\{  \bm{\Gamma},\bm{R}  \} ] \right)^2  }_{  \{  \bm{\Gamma}, \bm{R}  \}  }  -
  \bracket{    W[\{ \bm{\Gamma},\bm{R} \} ]}^2_{  \{  \bm{\Gamma}, \bm{R}  \}  } \nonumber \\
  &+& \frac{c}{2} \bracket{\bracket{   \left(  \oint_\gamma\frac{dz}{2\pi i} \text{Tr} \left[ \bm{G}(z) \bm{H}(z) \right] \right)^2}_{\bm{J}} }_{  \{  \bm{\Gamma}, \bm{R}  \} ,\{  \bm{\Gamma}^{\prime}, \bm{R}^{\prime}  \}  },
  \label{eq:Kumulant2}
\end{eqnarray}  
where the path probability $w[\{ \bm{\Gamma}, \bm{R} \}]$ appearing in $\kappa_1$ and $\kappa_2$ is calculated at $\mu=0$. Figure \ref{fig:firstCumulant} depicts the first two cumulants as a function of the radius $R$ defining a disk centered at $z=0$, for average connectivities $c=3$ and $c=10$. Each shaded region delimits the error involved in the numerical solution of Eq.~\eqref{eq:W} using the population dynamics algorithm \cite{Note1}. Figure \ref{fig:firstCumulant} compares our theoretical findings with results obtained from the exact numerical diagonalizations of $N \times N$ adjacency matrices $\bm{A}$ with different $N$. The diagonalization results for the second cumulant show a stronger dependence with the matrix dimension, but  they approach the theoretical results for increasing $N$. 

Since $\kappa_2$ is finite for $R > 0$, the variance of $\mathcal{N}_{\bm{A}}$ scales linearly with $N \gg 1$, akin to the weak repulsion between the eigenvalues of sparse random matrices \cite{Metz2015,metz2016large,castillo2018large}. This scaling behaviour is different from the complex Ginibre ensemble, where the dependence of  variance of $\mathcal{N}_{\bm{A}}$ with the system size shows three remarkably different regimes \cite{LacroixA,LacroixB}.  The second cumulant, shown in Fig. \ref{fig:firstCumulant}, displays a non-monotonic behavior with a maximum at a certain radius, whose location approaches  $R=1$ for increasing $c$, consistently with the sharp boundary of $\rho(x,y)$ in the dense limit $c \rightarrow \infty$ \cite{LacroixB,Tim2009}.  For small $c$, the first two cumulants converge to a finite value as $R \rightarrow 0^+$, due to the existence of a $\delta$-peak in $\rho(x,y)$ at $z=0$ \cite{Tim2009}. Thus, the theory allows to calculate the average and the variance of the weights characterizing the $\delta$-peak contributions to the eigenvalue distribution in the limit $N \rightarrow \infty$.

\begin{figure}[h]
\centering
\includegraphics[height=5cm,width=8cm]{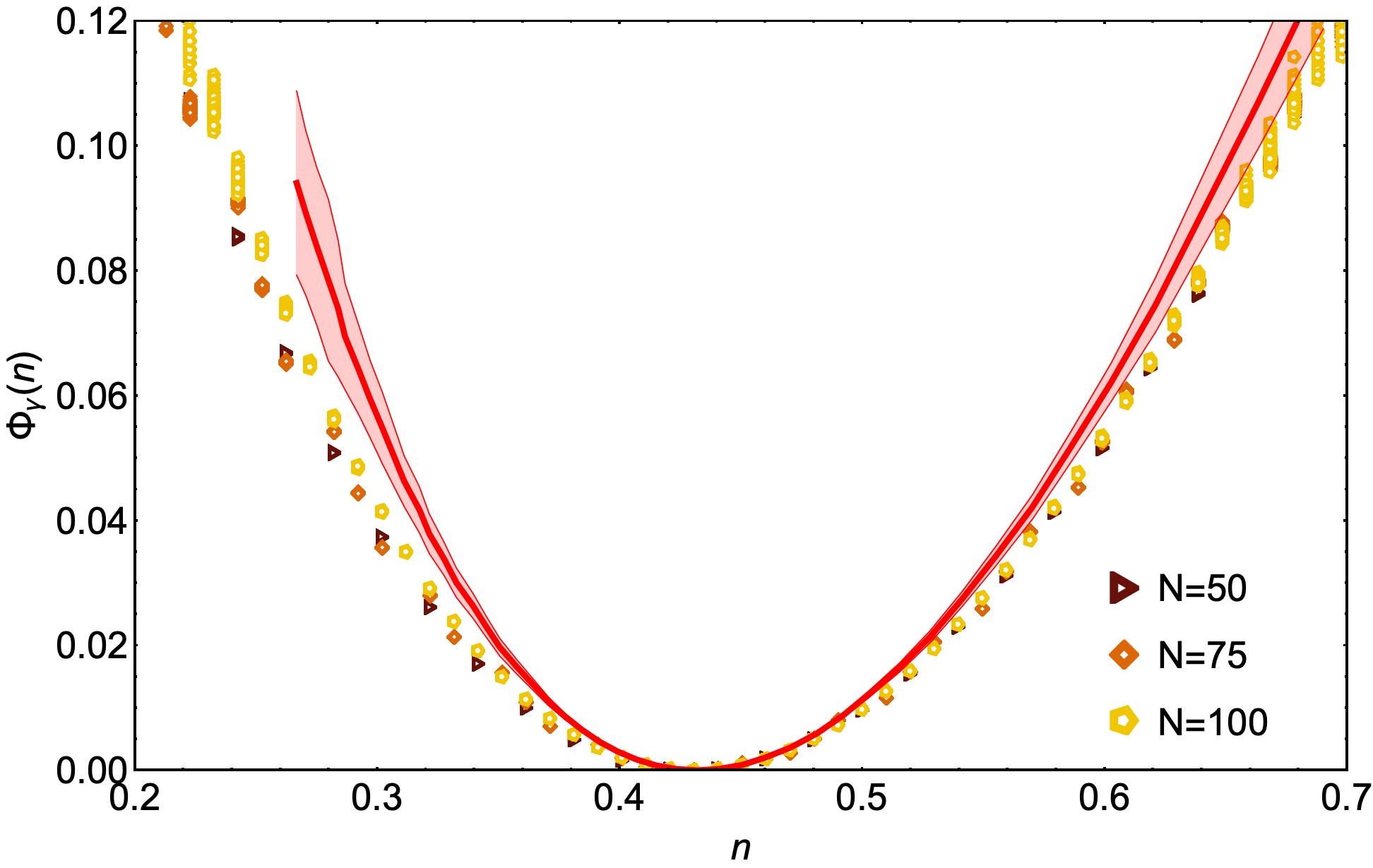}
\caption{Rate function $\Phi_\gamma(n)$ for the fraction $n$ of eigenvalues of $\bm{A}$ inside a disk of radius $0.5$ centered at the origin of the complex plane (see Eq.~(\ref{kalo})). The matrix $\bm{A}$ is the adjacency matrix of a random graph with mean connectivity $c=4$ and asymmetric couplings drawn from a Gaussian distribution with zero mean and variance $1/c$. The  red solid line corresponds to our theoretical findings for $N \rightarrow \infty$, while the markers are the results of numerical diagonalizations of $N \times N$ random matrices. The diagonalization results are averaged over $10^7$ samples and the process is repeated $10$ times, yielding the scatter plots in the figure.}
\label{fig:ratefunction}
\end{figure} 

In Fig. \ref{fig:ratefunction} we present the theoretical results and the direct diagonalization results for the  rate function $\Phi_\gamma(n)$ controlling the large deviations of the fraction $n=\frac{\mathcal{N}_{\bm{A}}}{N}$ of eigenvalues inside a disk of radius $R=0.5$. The shaded area in Fig. \ref{fig:ratefunction} bounds the error involved in the numerical
solution of Eq.~\eqref{eq:W}. The direct diagonalization results in Fig. \ref{fig:ratefunction} consistently approach the theoretical curve for increasing $N$, supporting the exactness of our theory. A striking property is the asymmetry of $\Phi_\gamma(n)$ around its minimum, located at $n = \kappa_1$. Sparse and asymmetric random matrices normally contain delocalized eigenvectors around $z=0$ and localized eigenvectors close to the boundary of $\rho(x,y)$ \cite{Amir2016,Zhang2019}. Since the eigenvalue repulsion is stronger within the delocalized region \cite{Zhang2019}, large fluctuations of $n$ corresponding to an attraction of more eigenvalues to inside the disk are less likely, resulting in a rate function that grows faster for $n > \kappa_1$ in comparison to $n < \kappa_1$. This property is at variance with the Ginibre ensemble \cite{Allez2014,LacroixB}, whose rate function $\Phi_\gamma(n)$ is symmetric around its minimum due to the absence of localized eigenvectors.

\paragraph{Conclusions.} While in the last  decades there has been a leap forward in understanding the statistical properties related to the spectrum of Hermitian random matrices, similar studies for non-Hermitian matrices are still in their infancy. This is mostly due to a lack of mathematical tools to analyse systems with asymmetric interactions. In this Letter, we have developed a powerful technique to study the typical and atypical eigenvalue fluctuations of infinitely large non-Hermitian random matrices $\bm{A}$.

We have presented a theory for the statistics of the number $\mathcal{N}_{\bm{A}}(\gamma)$ of eigenvalues within an arbitrary contour $\gamma \in \mathbb{C}$. The method does not rely on the analytic knowledge of the joint probability distribution of eigenvalues and it can be applied to various random-matrix ensembles, beyond the standard Gaussian ensembles of random matrix theory. In fact, we have formulated the theory for an arbitrary ensemble of non-hermitian random matrices, but we have derived explicit results for an ensemble of weighted random graphs with asymmetric couplings \cite{NewmanBook} . The main outcome is an effective theory for the cumulant generating function of $\mathcal{N}_{\bm{A}}(\gamma)$, from which we computed the first two cumulants of $\mathcal{N}_{\bm{A}}$ and its large deviation behaviour. In particular, we found that the large deviation probability of $\mathcal{N}_{\bm{A}}$ is asymmetric around its minimum, due to the existence of both delocalized and localized eigenvectors in the spectra of sparse asymmetric random matrices.

The generality of our approach opens the door to investigate the fluctuations of other observables describing the spectra of directed random networks, such as the fraction of real eigenvalues, the index, and the spectral radius. All these quantities play an important role to characterize the stability of large biological systems \cite{McMurtrie,allesina2012,Allez2014,Metz2019r}.

\acknowledgements{
I. P. C and F. L. M. thank London Mathematical Laboratory for financial support. F. L. M. also acknowledges a fellowship from CNPq/Brazil.
}

\bibliography{Bibliography.bib}

\end{document}